\documentclass[twocolumn,english,9pt,journal,letterpaper]{IEEEtran}
\usepackage[T1]{fontenc}
\usepackage{color}
\usepackage{babel}
\usepackage{array}
\usepackage{float}
\usepackage{multirow}
\usepackage{amsmath}
\usepackage{graphicx}
\usepackage[none]{hyphenat}
\usepackage{orcidlink}
\usepackage{amssymb}
\usepackage{mathtools}
\usepackage{subfigure}
\usepackage{cleveref}
\crefname{figure}{Fig.}{Fig.}
\crefname{equation}{}{}
\crefname{table}{Table}{Table}
\crefname{section}{Section}{Section}

\usepackage{cite}



\newcommand{\bs}[1]{\boldsymbol{#1}}

\begin{document}

\title{AC/DC Frequency-Dependent Jacobian: \\
Quantifying Grid Support and Stability Implications}

\author{
Dongyeong~Lee~\orcidlink{0000-0002-6570-6464}, Eros~Avdiaj~\orcidlink{0000-0003-1802-3342}, and Jef~Beerten~\orcidlink{0000-0002-8756-2983}

}

\maketitle

\begin{abstract}
This letter proposes an AC/DC frequency-dependent Jacobian analysis to identify the system support capabilities. 
In addition, the analyses reveal that stronger system support does not necessarily improve stability margins, suggesting that narrow-frequency-band (e.g., 4-40Hz) and AC-side-focused technical requirements may not yield the intended grid-forming stability contributions.
\end{abstract}
\section{Introduction}
Power system operators and research organizations have intensified efforts to standardize technical requirements for grid-forming (GFM) controls \cite{NESO_GFM_Practice_Guide_GB, AEMO_GFM_Aus, NERC_GFM_USA, ENTSOE_GFM_EU}.
Among these efforts, active and reactive power response-based transfer functions have been utilized to characterize the system support dynamics of GFM \cite{Shah}.
Their effectiveness for system stability analysis has been demonstrated \cite{Shah, FD-JB_Dey}.
There have also been technical suggestions to utilize these transfer functions as performance metrics of GFM, focusing on the AC side within the bandwidth of 4 to 40Hz \cite{ESIG_Test}.
In this letter, the combination of these active and reactive power response-based transfer functions is referred to as the frequency-dependent Jacobian (FD-JB), as they are a natural extension of the power-flow Jacobian formulation.
Although these approaches provide valuable insights, they are confined to AC-side dynamics and to the restricted frequency band where grid-support dynamics dominate. 
A lack of proper consideration of DC-side dynamics and a broader frequency range to account for the converter's dynamic behavior can result in an incomplete—or even misleading—assessment of a GFM unit’s ability to contribute to overall system operation and stability. 
Furthermore, the justification remains insufficient to show whether the identified supports stabilize the system.
To address these gaps, this letter proposes an extended analysis framework to assess system support characteristics, integration stability, and the system-stabilizing effects of GFM converters. The key contributions are summarized as follows:
\begin{itemize}
\item Detailed derivation of an extended AC/DC FD-JB model, representing system supportive characteristics and establishment of a formal bridging between admittance and FD-JB models.
\item Proposal of a unified analysis framework for systematic evaluation of system operation support capabilities, integration stability, and system stabilizing effects with the device.
\item Identification of critical limitations in technical requirements that focus primarily on AC side operation support via a limited band, which do not result in system stabilizing effects.
\end{itemize}
\begin{figure*}[!t]
\centering
\includegraphics[width=5.5in]{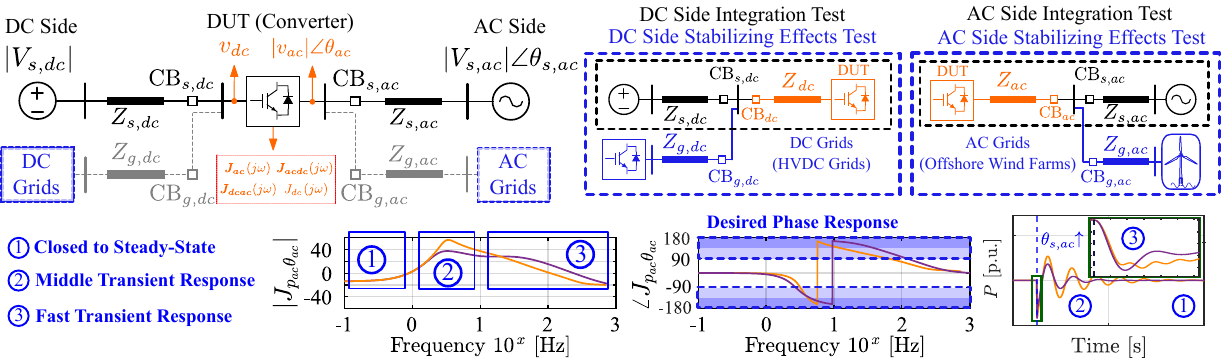}
\caption{Unified analysis framework with a benchmark test system to test the system support capabilities, system integration stability, and stabilizing effects.}
\label{fig: Benchmark test system}
\end{figure*}
\section{AC/DC Frequency Dependent Jacobian}
This section provides the derivation of the extended AC/DC FD-JB, including the bridging between admittance and FD-JB $\bs{J}(j\omega)$.
From instantaneous AC-side active power $P_{ac}$,  reactive $Q_{ac}$,  and DC-side power $P_{dc}$ of the device can be expressed based on 
$q$-axis leading and $d$-axis oriented $dq$-frame with respect to small perturbation $\Delta$ as
\begin{align}
    \Delta P_{ac}(s)&=v_{d,0}\Delta i_d + v_{q,0}\Delta i_{q}+  i_{d,0}\Delta v_{d} + i_{q,0}\Delta v_{q}
    \label{eq: P_ac small}\\
    \Delta Q_{ac}(s)&=v_{q,0}\Delta i_{d}-v_{d,0}\Delta i_q - i_{q,0}\Delta v_d + i_{d,0}\Delta v_q
    \label{eq: Q_ac small}\\
    \Delta P_{dc}(s)&=v_{dc,0}\Delta i_{dc} + i_{dc,0}\Delta v_{dc}
    \label{eq: P_dc small}
\end{align}
where subscript $0$ denotes the initial operating condition.
From \cref{eq: P_ac small,eq: Q_ac small,eq: P_dc small}, the AC/DC admittance representation can be found as
\begin{equation}
        \Delta\boldsymbol{i_{dq,dc}}(j\omega)=\boldsymbol{Y}(j\omega)\Delta\boldsymbol{v_{dq,dc}}(j\omega)
        \label{eq: dqdc currents admittance voltages}
\end{equation}
where $\boldsymbol{i_{dq,dc}}(j\omega)=[i_d(j\omega), i_q(j\omega), i_{dc}(j\omega)]^T$,  $\boldsymbol{v_{dq,dc}}(j\omega)=[v_d(j\omega), v_q(j\omega), v_{dc}(j\omega)]^T$ which are the components of $d$- and $q$-axis together with DC side of voltage and current.
$\boldsymbol{Y}(j\omega)$ is AC/DC admittance.
$v_d,v_q$ are $dq$ voltages, $i_d,i_q$ are $dq$ currents, and $v_{dc},i_{dc}$ are DC voltage and current.
In this paper, all the power directions are assumed to flow out from the converter to the interconnected systems.
The bold symbol denotes vectors or matrices. $|v_{ac}|, \theta_{ac}$ denote the AC-side voltage magnitude and phase angle of the device under the test (DUT).
With $\bs{Y}$ representing frequency-dependent admittance, FD-JB $\bs{J}(j\omega)$ enables the evaluation of device dynamic characteristics at frequency $\omega$ with the power system operation aspects ($P_{ac}, Q_{ac}, P_{dc}$ with respect to $|v_{ac}|,\theta_{ac}, v_{dc}$), which can be expressed as
\begin{align}
    \bs{J}(j\omega)
    &=\begin{bmatrix}
        \underbrace{\bs{J_{ac}}(j\omega)}_{2\times2} & \underbrace{\bs{J_{acdc}}(j\omega)}_{2\times1}\\
        \underbrace{\bs{J_{dcac}}(j\omega)}_{1\times2} & J_{dc}(j\omega)
    \end{bmatrix}
\end{align}
For the sake of conciseness, ($j\omega$) is omitted hereafter.
To underpin the effectiveness and modeling approach of the FD-JB, the derivation of $\mathbf{J}$ is presented.
To do this, $\bs{J}$ is mathematically bridged with $\bs{Y}$, while enabling efficient calculation of $\bs{J}$.
To do so, the interconnecting side voltage and its phase angle, which are considered as disturbances to the device, are expressed as
\begin{equation}
    |v_{ac}|=\sqrt{v_{d}^2+v_q^2},\quad     \theta_{ac}=\tan^{-1}(\frac{v_q}{v_d})
\end{equation}
where $v_{d}=|v_{ac}|\cos\theta_{ac}, v_q=|v_{ac}|\sin\theta_{ac}$.
$v_d=|v_{ac}|_0+\Delta|v_{ac}|$ and
$v_q=|v_{ac}|_0\Delta\theta_{ac}$.
Then, $\Delta\bs{v_{dq,dc}}$ and $\Delta\bs{i_{dq,dc}}$ in \cref{eq: P_ac small,eq: Q_ac small,eq: P_dc small} are replaced by \cref{eq: dqdc currents admittance voltages}.
Accordingly, $\bs{J}$ can be found as
\begin{equation}
\begin{aligned}
\Delta P_{\mathrm{ac}}
&= v_{d,0}\big(Y_{dd}\Delta v_{d} + Y_{dq}\Delta v_{q} + Y_{d,dc}\Delta v_{\mathrm{dc}}\big)+ i_{d,0}\Delta v_{d} \\
 &+ v_{q,0}\big(Y_{qd}\Delta v_{d} + Y_{qq}\Delta v_{q} + Y_{q,dc}\Delta v_{\mathrm{dc}}\big)+ i_{q,0}\Delta v_{q}\\
&=
\underbrace{(v_{d,0}Y_{dd} + v_{q,0}Y_{qd} + i_{d,0})}_{J_{p_{ac}v_{ac}}}\,\Delta |v_{ac}|\\
&+ \underbrace{(v_{d,0}Y_{dq} + v_{q,0}Y_{qq} + i_{q,0})}_{J_{p_{ac}\theta_{ac}}}\,\Delta \theta_{ac} \\
&\quad
+ \underbrace{(v_{d,0}Y_{d,dc} + v_{q,0}Y_{q,dc})}_{J_{p_{ac}v_{dc}}}\,\Delta v_{\mathrm{dc}}
\end{aligned}
\end{equation}
\begin{equation}
\begin{aligned}
\Delta Q_{\mathrm{ac}}
&= -v_{d,0}\big(Y_{qd}\Delta v_{d} + Y_{qq}\Delta v_{q} + Y_{q,dc}\Delta v_{\mathrm{dc}}\big)- i_{q,0}\Delta v_{d}\\
&  + v_{q,0}\big(Y_{dd}\Delta v_{d} + Y_{dq}\Delta v_{q} + Y_{d,dc}\Delta v_{\mathrm{dc}}\big)+ i_{d,0}\Delta v_{q}
\\ &=
\underbrace{(-v_{d,0}Y_{qd} + v_{q,0}Y_{dd} - i_{q,0})}_{J_{q_{ac}v_{ac}}}\,\Delta |v_{ac}|\\
& + \underbrace{(-v_{d,0}Y_{qq} + v_{q,0}Y_{dq} + i_{d,0})}_{J_{q_{ac}\theta_{ac}}}\,\Delta \theta_{ac} \\
&
+ \underbrace{(-v_{d,0}Y_{q,dc} + v_{q,0}Y_{d,dc})}_{J_{q_{ac}v_{dc}}}\,\Delta v_{\mathrm{dc}}
\end{aligned}
\end{equation}
\begin{equation}
\begin{aligned}
\Delta P_{\mathrm{dc}}
&= v_{\mathrm{dc},0}\big(Y_{dc,d}\Delta v_{d} + Y_{dc,q}\Delta v_{q} + Y_{dc,dc}\Delta v_{\mathrm{dc}}\big)+ i_{\mathrm{dc},0}\Delta v_{\mathrm{dc}}\\
&=
\underbrace{(v_{\mathrm{dc},0}Y_{dc,d})}_{J_{p_{dc}v_{ac}}}\,\Delta |v_{ac}|
+ \underbrace{(v_{\mathrm{dc},0}Y_{dc,q})}_{J_{p_{dc}\theta_{ac}}}\,\Delta \theta_{ac}\\
&+ \underbrace{(v_{\mathrm{dc},0}Y_{dc,dc} + i_{\mathrm{dc},0})}_{J_{p_{dc}v_{dc}}}\,\Delta v_{\mathrm{dc}}
\end{aligned}
\end{equation}
\begin{align}
\begin{bmatrix}
\Delta P_{\mathrm{ac}} \\[3pt]
\Delta Q_{\mathrm{ac}} \\[3pt]
\Delta P_{\mathrm{dc}}
\end{bmatrix}
&=
\underbrace{\
\begin{bmatrix}
J_{p_{ac}v_{ac}} & J_{p_{ac}v_{ac}} & J_{p_{ac}v_{ac}} \\[3pt]
J_{p_{ac}\theta_{ac}} & J_{q_{ac}\theta_{ac}} & J_{q_{ac}\theta_{ac}} \\[3pt]
J_{p_{dc}v_{dc}} & J_{p_{dc}v_{dc}} & J_{p_{dc}v_{dc}}
\end{bmatrix}
}_{\bs{J}}
\begin{bmatrix}
        \Delta |v_{ac}|\\
        \Delta \theta_{ac} \\
        \Delta v_{dc}
    \end{bmatrix}
\end{align}
As a result, $\bs{J}$ quantifies the device's system support capabilities, especially with respect to the changes in $|v_{ac}|, \theta_{ac},$ and $v_{dc}$ of its interconnecting points.
The bridge between $\bs{J}$ and $\bs{Y}$ implies $\bs{J}$ also inherits the device's dynamic influences on the system stability as $\bs{Y}$ does, while providing a power system-friendly interpretation.

\section{Unified Analysis Framework}
In this section, the proposed unified analysis framework is illustrated,
including how to interpret the results of $\bs{J}$ and how to test system supporting capabilities, system integration stability, and stabilizing effects.
Its graphical illustration is presented in \cref{fig: Benchmark test system}. 
\subsection{Benchmark Test System}
The test system models each AC/DC side of DUT as a voltage source with Thevenin impedance $Z_s$, connected to AC/DC grids through $Z_g$.
This structure enables a unified and flexible evaluation of system support characteristics, system integration stability, and the stabilizing effects of the DUT by circuit breakers (CBs) and impedances.
System support characteristics can be assessed by perturbing voltages, $|V_{s,ac}|\angle\theta_{s,ac}$ or $V_{s,dc}$, while $Z_{s}=0$ and opening $\text{CB}_{g}$.
More realistic interconnected conditions can be examined by introducing a non-zero $Z_{s}$, allowing analysis of system integration stability with interactions between DUT and AC/DC systems.
Moreover, subsystem-level studies (e.g., microgrid) can be performed by opening $\text{CB}_s$ and closing $\text{CB}_g$ while connecting considered sub-systems on AC/DC grids with $Z_g$.
Finally, closing all CBs allows assessment of DUT's system stabilizing effects on the complete AC/DC system.
Overall, it provides a flexible framework that supports the interoperability-oriented design of AC/DC grids.
\subsection{Interpretation of FD-JB and Ideal Behavior in Frequency-Domain}
FD-JB in the bottom part of \cref{fig: Benchmark test system} shows three sections of the frequency region with the examples of $J_{p_{ac}\theta_{ac}}$ of two cases (purple and orange colors) and time-domain response results when it is tested via $Z_s=0$ and all $\text{CB}_g$ opened.
The low-frequency band represents the very slow response, which is close to steady-state.
The middle frequency band represents a relatively slow transient response, while the high frequency band represents a relatively fast transient response.
These middle and high frequency bands can present expected GFM's system operational support (e.g., 4-40Hz in \cite{ESIG_Test}).
For the magnitude, a higher value denotes a stronger response from the DUT.
The phase information of $\bs{J}$ is also important to identify the intended grid support as it is related to the response direction with respect to the system changes, particularly, considering $J_{p_{ac}\theta_{ac}}, J_{q_{ac}{v_{ac}}},$ and $J_{p_{dc}v_{dc}}$ to support power system operation.
Therefore, the desired GFM's dynamic responses for system operation support are a larger magnitude and a phase angle closer to 180$^\circ$, which reflect responses of GFM by resisting the changes of the system. 
In the example of $J_{p_{ac}\theta_{ac}}$ in \cref{fig: Benchmark test system}, the purple line shows a higher magnitude in the high frequency band, a lower magnitude in the middle frequency, and the same one in the low frequency compared to the orange line.
Time-domain responses show alignment with $J_{p_{ac}\theta_{ac}}$.
According to $J_{p_{ac}\theta_{ac}}$, the purple line shows a stronger response in the initial transient but a weaker late transient, eventually showing a similar tendency compared to the orange line as they get closer to the steady state.
However, even though it demonstrates its effectiveness in validating the system operation support aspect, further investigation is required to verify its implications for system stability.
\section{Case Studies}
In this section, case studies are presented to investigate the implications of FD-JB on the system stability.
The used DUT is a modular multilevel converter.
The control scheme is virtual synchronous machine for GFM with virtual inductance $L_v$ or grid-following (GFL) with DC voltage droop between $v_{dc}$ and $p_{ac}$ as $e_p=K_{d,dc}(v_{dc}-v^*_{dc})+p^*_{ac}-p_{ac}$. 
$e$ and superscript * denote the error and reference signals for control loops with droop gain $K_{d,dc}$. 
\subsection{System Support Capabilities \& System Integration Stability}
\begin{figure}[!t]
\centering
\includegraphics[width=3.0in]{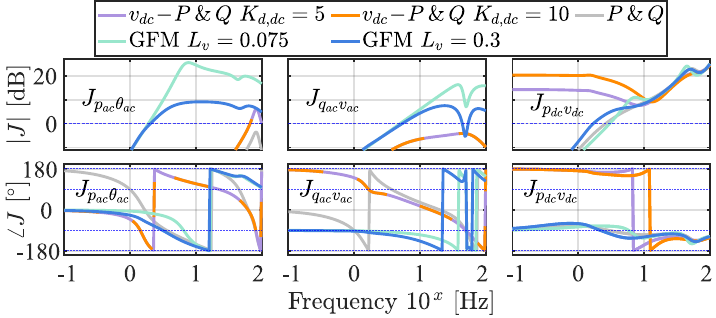}
\caption{FD-JB of $J_{p_{ac}\theta_{ac}}, J_{q_{ac}v_{ac}},$ and $J_{p_{dc}v_{dc}}$ with different converters.}
\label{fig: ACDC_JBs}
\end{figure}

Using the developed test system, system-supportive characteristics are tested to verify the effectiveness of the proposed extended AC/DC $\bs{J}$ by opening all $\text{CB}_{g}$.
\cref{fig: ACDC_JBs} shows three key elements of $\bs{J}$: $J_{p_{ac}\theta_{ac}}, J_{q_{ac}v_{ac}},$ and $J_{p_{dc}v_{dc}}$. 
Here, $P\&Q$ case refers to the case of GFL with $K_{d,dc}=0$. 
$P_{ac,0}=1.0$p.u, is assumed. 

\begin{figure}[!t]
\centering
\includegraphics[width=3.0in]{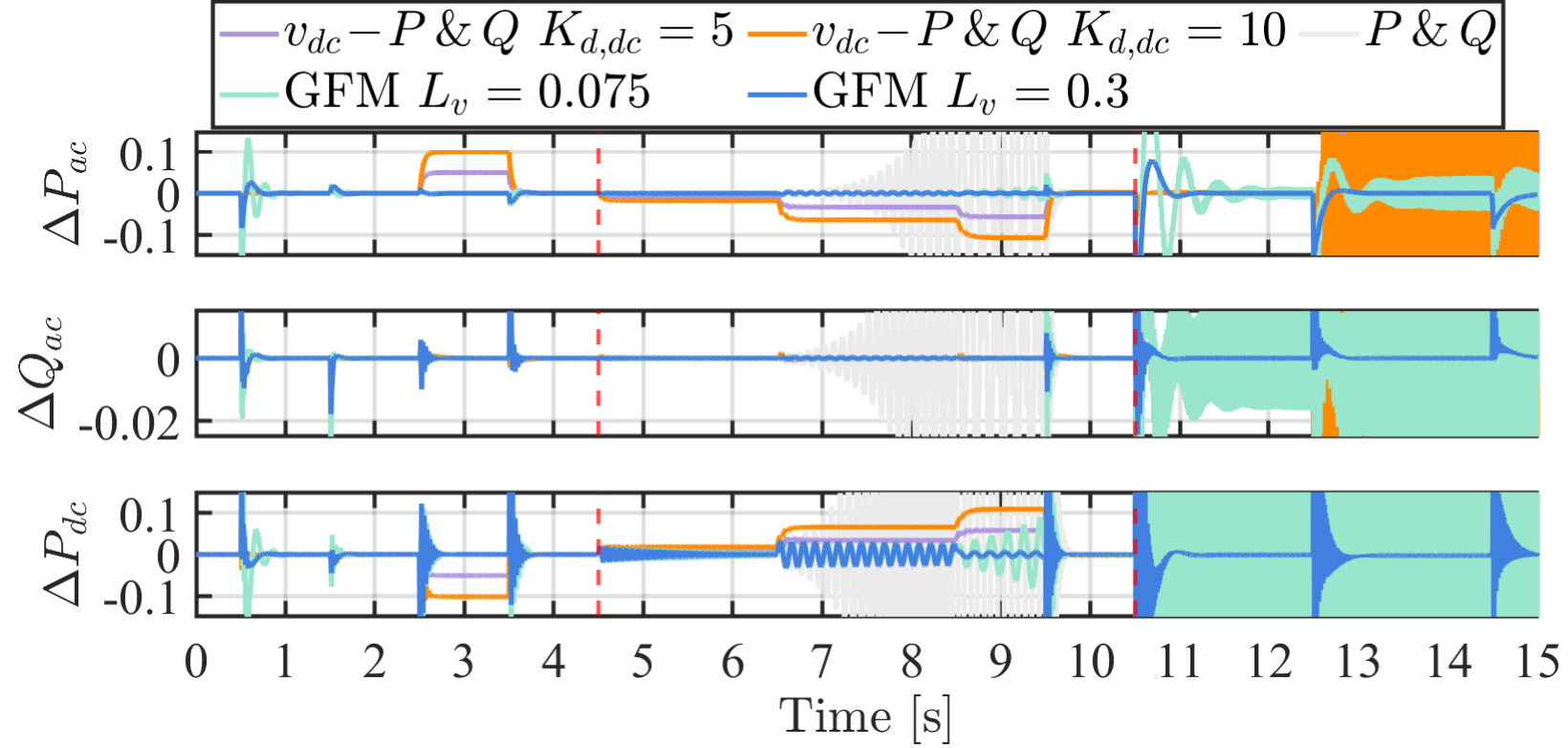}
\caption{EMT results of system support and weak-grid integration tests.}
\label{fig: ACDC Test system results}
\end{figure}

In the 0-3.5s interval of \cref{fig: ACDC Test system results}, $\Delta\theta_{s,ac}=2^\circ, \Delta |V_{s,ac}|=0.01$p.u., and $\Delta V_{s,dc}=0.01$p.u. are sequentially applied and restored at 3.5s. 
It shows the responses of DUT corresponding to \cref{fig: ACDC_JBs}.
GFM exhibits stronger responses to AC-side disturbances at 0.5s and 1.5s, while inducing a significant $\Delta P_{dc}$.
This indicates that, without proper consideration of DC-side dynamics, the expected grid support of GFM might deviate from AC-side-only considerations.
GFLs with $K_{d,dc}$ show significant $\Delta P_{dc}$ and $\Delta P_{ac}$, including a steady-state response according to droop control, i.e., due to the increases in $v_{dc}$. 
These observations confirm that $\bs{J}$ effectively captures the operational grid support characteristics.
In the 4.5-9.5s interval of \cref{fig: ACDC Test system results}, DC weak-grid stability is tested by sequentially increasing $Z_{s,dc}=[0.001+j0.15, 0.0035+j0.525, 0.006+j0.9]$ p.u. and restoring it at 9.5s. 
GFL $K_{d,dc}=0$ becomes unstable, whereas GFLs with $K_{d,dc}=5,10$ remain stable even as the grid weakens.
Notably, GFM with a lower $L_v$ exhibits instability, while the one with a higher $L_v$ does not.
After 10.5s in \cref{fig: ACDC Test system results}, AC weak-grid stability is tested by sequentially increasing $Z_{s,ac}=[j0.325, j0.425, j0.475]$ p.u.
Only the GFM with the higher $L_v$ remains stable, while others become unstable.
Thus, although $L_v=0.075$ yields stronger AC-side support, it does not remain stable under weak-grid conditions.
Hence, stronger grid support characteristics do not guarantee better stability.

The results demonstrate the effectiveness of the test system in evaluating system operational support and system integration stability.
However, stronger support reflected in $\bs{J}$ does not guarantee improved stability, leaving its contribution to overall system stabilization unclear.
To clarify this, AC/DC sides are tested individually in the following subsections.
For each test, the subsystem is considered with GFL $P\&Q$ by closing $\text{CB}_{g}$ with $|Z_g|$= 0.01 p.u., whereas the non-tested side is represented by an ideal voltage source.
\subsection{Stabilizing Effects Analysis with DC Weak Grid}
\label{subsec:DC_Stabilizing}
\begin{figure}[!t]
\centering
\includegraphics[width=3.0in]{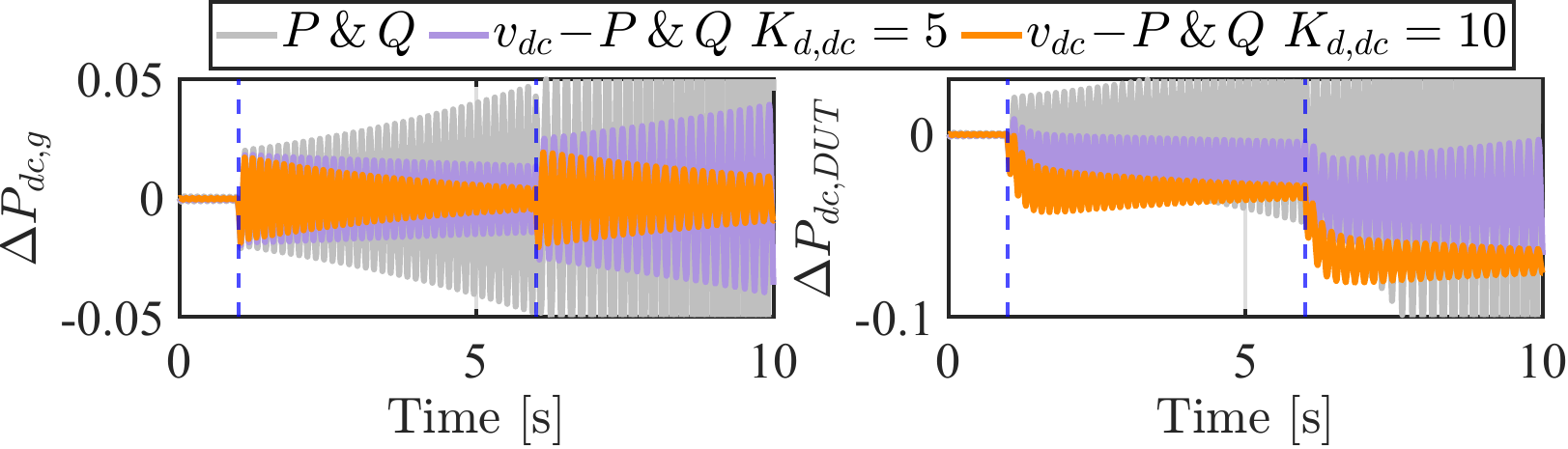}
\caption{EMT results of DC side stabilizing effects where DC weak-grid with $|Z_{s,dc}|$ increases at 1s and 6s.}
\label{fig: DC Stabilizing effects test}
\end{figure}

The DC side stabilizing effects test results are shown in \cref{fig: DC Stabilizing effects test}.
$Z_{s,dc}=[0.0018+j0.27, 0.0024+j0.36]$ in p.u. are sequentially applied at 1s and 6s.
The results show that, although the $K_{d,dc}=5$ case remains stable in the weak grid integration test without subsystem consideration in \cref{fig: ACDC Test system results}, it fails to stabilize the system when the subsystem is included.
In contrast, $K_{d,dc}=10$ shows stable results with the subsystem.
It implies that a superior $\bs{J}$ might lead to stronger stabilizing effects as long as its own stability is guaranteed.
\subsection{Stabilizing Effects Analysis with AC Weak Grid}
\begin{figure}[!t]
\centering
\includegraphics[width=3.0in]{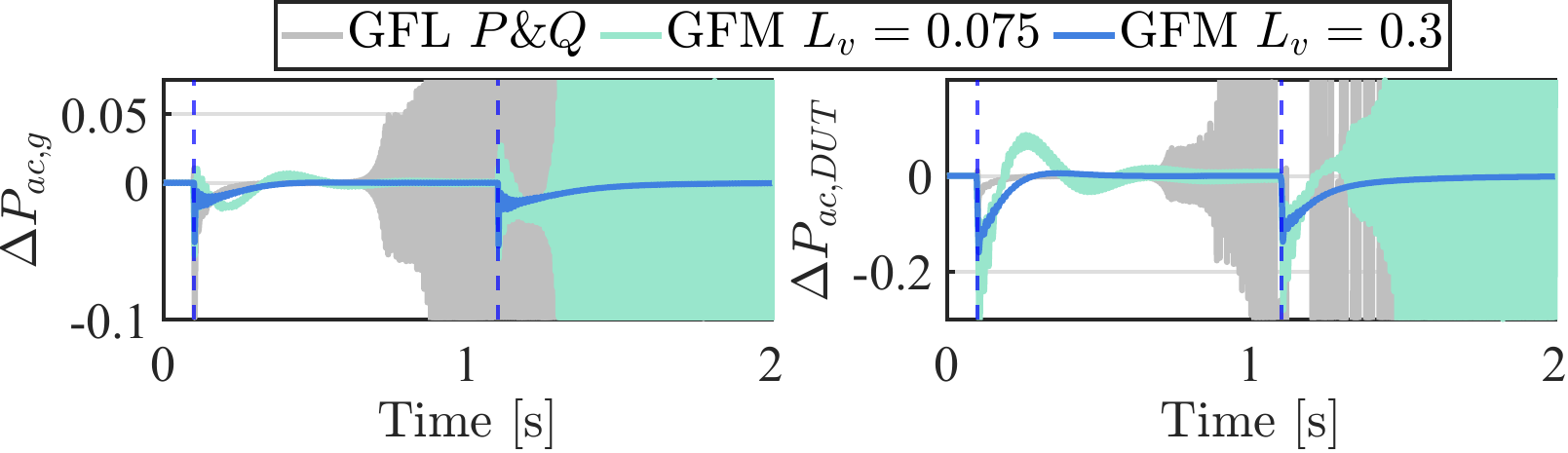}
\caption{EMT results of AC side stabilizing effects where AC-weak grid with $|Z_{s,ac}|$ increases at 1s and 6s.}
\label{fig: AC Stabilizing effects test}
\end{figure}

The AC side stabilizing effects test results are shown in \cref{fig: AC Stabilizing effects test}.
$Z_{s,ac}=[j0.22, j0.245]$ in p.u. are sequentially applied at 1s and 6s.
As the low-frequency bands of $J_{p_{ac}\theta_{ac}}$ and $J_{q_{ac}v_{ac}}$ in \cref{fig: ACDC_JBs} are small and similar, GFMs do not show significant differences when approaching steady-state conditions. 
In the middle- and high-frequency bands of FD-JB, the lower $L_v$ case shows a higher magnitude, indicating stronger support as discussed earlier.
This is reflected in its stronger instantaneous response between 0.1s and 1.1s, consistent with \cite{ESIG_Test}.
At 1.1s, despite its stronger system support characteristics, the lower $L_v$ case becomes unstable as the system weakens.
If only considering FD-JB, the lower $L_v$ case would be expected to yield stronger stabilizing effects, as in \cref{subsec:DC_Stabilizing}. 
However, consistent with the earlier findings that stronger support does not ensure better stability, this case shows that it also does not ensure stabilizing effects, since it becomes unstable under weak-grid conditions.
As a result, strong grid support does not necessarily lead to strong stabilizing effects.
Therefore, it is not enough to consider only system support characteristics (4-40Hz) from FD-JB to fulfill the expectations of GFM for improving the system stability.

\section{Conclusions}
This letter proposes the AC/DC FD-JB framework and demonstrates its effectiveness through a dedicated test system. The results reveal that AC-side-focused and band-limited assessments of GFM behavior can be misleading: enhanced system-support characteristics do not necessarily imply better stability or stronger stabilizing effects. Moreover, whether a GFM delivers the expected stabilizing contribution also depends on its own stability under relevant grid conditions. Thus, AC/DC FD-JB offers useful insight into system-support characteristics, but it must be complemented by holistic stability studies for the reliable design of AC/DC grids.

\bibliographystyle{IEEEtran}
\bibliography{ref}

\vfill
\end{document}